\documentclass[ssuperscriptaddress,nofootinbib,tightenlines,preprintnumbers]{revtex4}

\usepackage{amsfonts}
\usepackage{multirow}
\usepackage{mathrsfs}
\usepackage{graphicx}
\usepackage{amsmath}
\usepackage{amssymb}
\usepackage{bm}
\usepackage{bbm}
\usepackage{color}
\usepackage{slashed}
\usepackage{diagbox}
\usepackage{ulem}
\usepackage{slashbox}
\usepackage{wrapfig}
\usepackage{hyperref}
\usepackage{cleveref}
\usepackage{textpos}
\numberwithin{equation}{section}

\begin{document}
\title{\mbox{}\\[10pt]
Reduction of Feynman Integrals in the Parametric Representation}

\preprint{Alberta Thy 2-19}


\author{Wen Chen~\footnote{wchen1@ualberta.ca}}
\affiliation{Department of Physics, University of Alberta, Edmonton, AB T6G 2E1, Canada\vspace{0.2cm}}

\date{\today}

\begin{abstract}

In this paper, the reduction of Feynman integrals in the parametric representation is considered. This method proves to be more efficient than the integration-by-part (IBP) method in the momentum space. Tensor integrals can directly be parametrized without performing tensor reductions. The integrands of parametric integrals are functions of Lorentz scalars, instead of four momenta. The complexity of a calculation is determined by the number of propagators that are present rather than the number of all the linearly independent propagators. Furthermore, the symmetries of Feynman integrals under permutations of indices are transparent in the parametric representation. Since all the indices of the propagators are nonnegative, an algorithm to solve those identities can easily be developed, which can be used for automatic calculations.

\end{abstract}

\maketitle

\section{Introduction}

Nowadays high precision calculations are more and more important in high energy physics, which rely on calculations of multi-loop Feynman integrals. A standard strategy to calculate Feynman integrals is to reduce a large number of integrals to a smaller set of master integrals and then calculate the master integrals either analytically or numerically. A widely used technique to reduce Feynman integrals is the integration-by-part (IBP) method \cite{Tkachov:1981wb,Chetyrkin:1981qh}. Combining with various algorithms to solve IBP identities \cite{Baikov:1996iu,Tarasov:1998nx,Laporta:2001dd,Smirnov:2005ky,Lee:2008tj,Kant:2013vta,vonManteuffel:2014ixa}, IBP method proves to be very powerful in practice. Many published programs that implement these algorithms can be found \cite{Anastasiou:2004vj,Smirnov:2008iw,Studerus:2009ye,Lee:2012cn,vonManteuffel:2016xki,Maierhoefer:2017hyi,Smirnov:2019qkx}.

Nevertheless, there are some disadvantages for the traditional IBP method. A Feynman integral is a function of Lorentz invariants, while the integrand is not. Thus the original Feynman integral contains redundant information, of which Lorentz-invariance identities \cite{Gehrmann:1999as} are direct consequences. In order to reduce the number of independent external momenta, one has to reduce tensor integrals into a set of scalar integrals before using the IBP technique. However, for tensors of high ranks, the reduction by itself is very complicated. Furthermore, in order to construct IBP identities, all the linearly independent propagators should be considered, though some of them may not be present in the integrals to be reduced.

These problems can be solved by considering Feynman integrals in the parametric representation. It was suggested in ref. \cite{Lee:2014tja} that IBP relations can be derived directly in the Lee-Pomeransky representation \cite{Lee:2013hzt}. In this paper, a similar representation is used. It is shown that linear relations between integrals in this representation can directly be constructed. Feynman integrals can be reduced to master integrals by solving these linear relations, just as in the traditional IBP method. The advantage of this method is that it facilitates the reduction of tensor integrals (To be discussed in a subsequent paper \cite{Chen:2019fzm}).

This paper is organized as follows. In \cref{sec:IdentFeynmIntParamRepr}, identities between Feynman integrals in the parametric representation are constructed. The detail derivation of these identities can be found in \cref{app:DerIdent}. In \cref{sec:Alg}, we present a naive algorithm to solve these identities.

\section{Identities between Feynman Integrals in the Parametric Representation}\label{sec:IdentFeynmIntParamRepr}

It is well-known that a dimensionally regularized $L$-loop Feynman integral with $n$ propagators can be parametrized by integrals of the following structure \cite{Feynman:1949zx,Nakanishi:1957aa,Symanzik:1958aa}:

\begin{equation}\label{eq:FeynmInt}
I=\frac{\Gamma(-\lambda_f)}{\prod_{i=1}^n\Gamma(\lambda_i+1)}\int d\Pi^{(n)}U^{\lambda_u}F^{\lambda_f}\prod_{i=1}^nx_i^{\lambda_i},\quad \lambda\notin\mathbb{Z}^-.
\end{equation}

\noindent Here $d\Pi^{(n)}\equiv dx_1dx_2\cdots dx_n\delta\left(\sum_i\left|x_i\right|-1\right)$, where the sum in the delta function runs over any nontrivial subset of $\{x_1,x_2,\cdots x_n\}$. $U$ and $F$ are homogeneous polynomials of $x$ of degrees $L$ and $L+1$ respectively. By virtue of Mellin transformation, we may rewrite the integral in \cref{eq:FeynmInt} in the following form:

\begin{equation*}
\begin{split}
I=&\frac{\Gamma(-\lambda_u-\lambda_f)}{\Gamma(-\lambda_u)\prod_{i=1}^n\Gamma(\lambda_i+1)}\int dx_{n+1}d\Pi^{(n)}(Ux_{n+1}+F)^{\lambda_u+\lambda_f}x_{n+1}^{-\lambda_u-1}\prod_{i=1}^nx_i^{\lambda_i}\\
\equiv&\frac{\Gamma(-\lambda_0)}{\prod_{i=1}^{n+1}\Gamma(\lambda_i+1)}\int dx_{n+1}d\Pi^{(n)}\mathcal{F}^{\lambda_0}\prod_{i=1}^{n+1}x_i^{\lambda_i},\quad \lambda\notin\mathbb{Z}^-.
\end{split}
\end{equation*}

\noindent It's easy to prove that the sum in the delta function in $d\Pi^{(n)}$ can be extended to include $x_{n+1}$. Thus we have

\begin{equation}\label{eq:FeynmInt2}
I=\frac{\Gamma(-\lambda_0)}{\prod_{i=1}^{n+1}\Gamma(\lambda_i+1)}\int d\Pi^{(n+1)}\mathcal{F}^{\lambda_0}\prod_{i=1}^{n+1}x_i^{\lambda_i}\equiv\int d\Pi^{(n+1)}\mathcal{I}^{(-n-1)},\quad \lambda\notin\mathbb{Z}^-,
\end{equation}

\noindent where $\mathcal{I}^{(-n-1)}$ is a homogeneous function of $x$ of degree $-n-1$.

Tensor integrals can be parametrized by the standard procedure by parametrizing the denominators by Gauss integrals, shifting the loop momenta, and replacing loop momenta in the numerator by a sum of products of metric tensors. Alternatively, they can be parametrized by the generator method developed in \cref{app:ParTensInt} (cf. \cref{eq:app:ParTensInt}).

According to the derivation in \cref{app:DerIdent}, we have the following identities:

\begin{equation}
0=\int d\Pi^{(n+1)}\frac{\partial}{\partial x_i}\mathcal{I}^{(-n)}+\delta_{\lambda_i0}\int d\Pi^{(n)}\left.\mathcal{I}^{(-n)}\right|_{x_i=0},\qquad i=1, 2,\cdots, n+1,~\lambda\notin\mathbb{Z}^-,\label{eq:IBP1}
\end{equation}

\noindent where $\delta_{\lambda_i0}$ is the Kronecker delta. This equation can be understood as a generalization of IBP identities in the parametric representation \cite{Lee:2013hzt,Lee:2014tja}. To see this, we choose $d\Pi^{(n)}=\prod_{i=1}^{n+1}dx_i\delta(x_{n+1}-1)=\prod_{i=1}^{n}dx_i$. Then \cref{eq:IBP1} becomes

\begin{equation}
\begin{split}
0=&\int\prod_{j=1}^{n}dx_j\frac{\partial}{\partial x_i}\left[\prod_{j=1}^{n}x_j^{\lambda_j}\left(\left.\mathcal{F}\right|_{x_{n+1}=1}\right)^{\lambda_0}\right]+\delta_{\lambda_i0}\int\prod_{j\neq i}dx_j\left(\left.\mathcal{F}\right|_{x_i=0,x_{n+1}=1}\right)^{\lambda_0}\prod_{j\neq i}x_j^{\lambda_j}\\
&\equiv\int\prod_{i=j}^{n}dx_j\frac{\partial}{\partial x_i}\left[\prod_{j=1}^{n}x_j^{\lambda_j}\mathcal{G}^{\lambda_0}\right]+\delta_{\lambda_i0}\int\prod_{j\neq i}dx_j\left(\left.\mathcal{G}\right|_{x_i=0}\right)^{\lambda_0}\prod_{j\neq i}x_j^{\lambda_j},\qquad i=1, 2,\cdots, n,~\lambda\in\mathbb{Z}^-,
\end{split}
\end{equation}

\noindent which are exactly the IBP identities in the Lee-Pomeransky representation(except for that here we don't consider the cases where $\lambda\notin\mathbb{Z}^-$, which is unnecessary in practice).

Among identities \cref{eq:IBP1}, some are recurrence relations between integrals with shifted dimensions \cite{Tarasov:1996br,Laporta:2001dd}. Linear relations free of dimensional recurrence can be derived by using the method of parametric annihilators \cite{Baikov:1996iu,Lee:2013hzt,Lee:2014tja,Bitoun:2017nre} or the syzygy-equation method \cite{Larsen:2015ped,Boehm:2017wjc,Boehm:2018fpv}.\footnote{The latter method is based on Baikov representation \cite{Baikov:1996iu}. Nevertheless, it can easily be applied to the representation used in this paper.} For our approach, we do need these dimensional-recurrence relations, because tensor integrals are parametrized by integrals with shifted dimensions.

By applying \cref{eq:IBP1}, integrals associated with subdiagrams may arise, some of which may be scaleless. Scaleless integrals can be identified by the criterion that equation

\begin{equation}
\sum_{i=1}^{n}k_ix_i\frac{\partial\mathcal{G}}{\partial x_i}=\mathcal{G}
\end{equation}

\noindent has a nontrivial $x$-independent solution for $k$ \cite{Lee:2013mka}.

\section{the Algorithm}\label{sec:Alg}

In this section, we give a brief description of the algorithm we use to solve the linear relations in \cref{eq:IBP1}. We use an algorithm similar to that in ref. \cite{Laporta:2001dd} combined with the application of symbolic rules \cite{Lee:2012cn}.

An ordering for the integrals is prescribed. Integrals of the highest priority is solved first. By fixing the values of $\lambda$, \cref{eq:IBP1} can be solved symbolically. These solutions play the role of symbolic rules. For one-loop integrals, these symbolic rules are complete, in the sense that any one-loop integral can be reduced to master integrals by applying these rules. Thus the reduction of one-loop integrals is extremely fast by using this algorithm. While for multi-loop integrals, these symbolic rules are incomplete. In this case, we have to reduce the unreduced integrals by solving \cref{eq:IBP1} with the explicit values of $\lambda$ substituted in. These identities can be solved by Gauss elimination.

For convenience, we may express a parametric integral by a standard Feynman integral defined in dimension $-2\lambda_0\equiv d+m$, where $d$ is the space-time dimension and $m$ is a nonnegative integer. This kind of integrals can be numerically evaluated by using FIESTA \cite{Smirnov:2008py}. Since \cref{eq:IBP1} contains dimensional-recurrence relations, in order the process to terminate, we use an explicit cut-off: $m\geq0$.

As a trivial example, we consider the reduction of the tadpole integral:

\begin{equation*}
\begin{split}
I(\lambda_0,\lambda_1)\equiv&(-1)^{\lambda_1}i\pi^{\lambda_0}\int d^{-2\lambda_0}l\frac{1}{(l^2-m^2)^{1+\lambda_1}}\\
=&\frac{\Gamma(-\lambda_0)}{\Gamma(\lambda_1+1)\Gamma(\lambda_2+1)}\int d\Pi^{(2)}(m^2x_1^2+x_1x_2)^{\lambda_0}x_1^{\lambda_1}x_2^{\lambda_2},
\end{split}
\end{equation*}

\noindent where $\lambda_2=-2\lambda_0-\lambda_1-2$. By applying \cref{eq:IBP1}, we get following linear relations:

\begin{align*}
0=&I(\lambda_0,\lambda_1)-(\lambda_1+1)I(\lambda_0-1,\lambda_1+1),\\
0=&(2\lambda_0+\lambda_1)I(\lambda_0-1,\lambda_1)-2m^2(\lambda_1+1)I(\lambda_0-1,\lambda_1+1)+I(\lambda_0,\lambda_1-1).
\end{align*}

\noindent Here it is understood that

\begin{equation*}
I(\lambda_0,-1)=\frac{\Gamma(-\lambda_0)}{\Gamma(\lambda_2+1)}\int d\Pi^{(1)}\left(\left.(m^2x_1^2+x_1x_2)\right|_{x_1=0}\right)^{\lambda_0}x_2^{\lambda_2}=0.
\end{equation*}

\noindent These linear relations can be solved symbolically. The solutions read

\begin{align*}
I(\lambda_0,\lambda_1)=&\frac{1}{\lambda_1}I(\lambda_0+1,\lambda_1-1),\quad\lambda_1\geq1,\\
I(\lambda_0,\lambda_1)=&\frac{2m^2}{2\lambda_0+\lambda_1+2}I(\lambda_0+1,\lambda_1)-\frac{1}{2\lambda_0+\lambda_1+2}I(\lambda_0+1,\lambda_1-1).
\end{align*}

\noindent These solutions play the role of symbolic rules. Obviously these rules are complete in this example. The first rule can be used to reduce the index $\lambda_1$, and the second rule can be used to reduce the spacetime dimension.

As a less trivial example, we consider the reduction of the following two-loop massless double-box integral:

\begin{equation*}
M^{\mu\nu}=\int d^dl_1d^dl_2P(-1,-1,-1,-1,-1,-1,-1)l_1^\mu l_2^\nu,
\end{equation*}

\noindent where

\begin{equation*}
P(i_1,i_2,i_3,i_4,i_5,i_6,i_7)\equiv l_1^{2i_1}l_2^{2i_2}(l_1+k_1)^{2i_3}(l_1-k_2)^{2i_4}(l_1+l_2+k_1)^{2i_5}(l_1+l_2-k_2)^{2i_6}(l_1+l_2-k_2-k_3)^{2i_7}.
\end{equation*}

\noindent We put $(k_1+k_2)^2=8$, and $(k_1+k_3)^2=-1$. By expressing parametric integrals in terms of standard Feynman integrals, the result reads:

\begin{equation*}
\begin{split}
M^{\mu\nu}=&\frac{1}{16d-58}\left[7 (d-3) k_1^{\mu } k_2^{\nu }+(d-4) k_2^{\mu } k_2^{\nu }+8 (d-4) k_3^{\mu } k_2^{\nu }+7 (d-11) k_1^{\mu } k_3^{\nu }+(d-11) k_2^{\mu } k_3^{\nu }\right.\\
&\left.+8 (d-11) k_3^{\mu } k_3^{\nu }-28 g^{\mu  \nu }-49 k_1^{\mu } k_1^{\nu }-56 k_3^{\mu } k_1^{\nu }\right]\int d^dl_1d^dl_2P(-1,-1,-1,-1,-1,-1,-1)\\
&+\frac{d-3}{8d-29}\left[\frac{1}{8} (3 d-14) g^{\mu  \nu }+\frac{7}{64} (1-2 d) k_1^{\mu } k_1^{\nu }+\frac{1}{64} (29-8 d) k_2^{\mu } k_1^{\nu }+\frac{1}{64} (-68d^2+10d+119) k_1^{\mu } k_2^{\nu }\right.\\
&+\frac{1}{448} (-62d^2+10d+91) k_2^{\mu } k_2^{\nu }+\frac{1}{8} (1-2 d) k_3^{\mu } k_1^{\nu }+\frac{1}{56} (-62d^2+10d+91) k_3^{\mu } k_2^{\nu }\\
&\left.+\frac{1}{32} (5d^2-38d+49) k_1^{\mu } k_3^{\nu }+\frac{1}{224} (5d^2-38d+49) k_2^{\mu } k_3^{\nu }+\frac{1}{28} (5d^2-38d+49) k_3^{\mu } k_3^{\nu }\right]\\
&\times\int d^{d+2}l_1d^{d+2}l_2P(-1,-1,-1,-1,-1,-1,-1)+\dots,
\end{split}
\end{equation*}

\noindent where the ellipsis represents contributions of subdiagrams, which are too complicated to be presented here. All the rest master integrals are

\begin{align*}
&\int d^dl_1d^dl_2P(-1, -1, 0, 0, 0, 0, -1),\\
&\int d^dl_1d^dl_2P(0, -1, -1, 0, 0, -1, 0),\\
&\int d^dl_1d^dl_2P(-1, -1, 0, 0, -1, -1, 0),\\
&\int d^dl_1d^dl_2P(0, 0, -1, -1, -1, -1, 0),\\
&\int d^dl_1d^dl_2P(-1, -1, -1, -1, 0, 0, -1),\\
&\int d^dl_1d^dl_2P(-1, -1, 0, -1, -1, 0, -1).
\end{align*}

\noindent The calculation is carried out by using a private code. To check the calculation, we contract the tensor integral with some external momenta, and reduce the resulting integral by using FIRE \cite{Smirnov:2019qkx}. The basis chosen by FIRE contains an integral with a double propagator. This integral can be further reduced by using our code. The final result for the tensor integral thus obtained is consistent with the one obtained by using our code. For integrals with simpler topologies, we have also verified the calculations numerically by using FIESTA \cite{Smirnov:2008py} (Notice that FIESTA can be used to evaluate integrals with numerators and with shifted dimensions).

\section{Discussion}

In this paper, the reduction of Feynman integrals in the parametric representation is considered. A representation similar to the Lee-Pomeransky representation is used. Tensor integrals can directly be parametrized by using a generator method. Identities between the parametric integrals are derived. Feynman integrals are reduced to master integrals by solving these identities. This method has many advantages over the traditional IBP technique, as is discussed at the very beginning of this paper. Symbolic rules can be derived out of these identities. One-loop integrals can be reduced to master integrals merely by applying these rules. For multi-loop integrals, these rules are incomplete. Thus we can not get rid of the Gauss elimination, which is less efficient. Though the symbolic rules are incomplete, the reduction is as complete as that of the traditional IBP method in the sense that numbers of master integrals obtained by these two methods are the same.

\vspace{0.2 cm}
\begin{acknowledgments}

The author thanks Shuai Liu for testing the code. This work was supported by the Natural Sciences and Engineering Research Council of Canada.

\end{acknowledgments}

\appendix

\section{Parametrization of Tensor Integrals}\label{app:ParTensInt}

It is well-known that a propagator $\frac{1}{D_i}\equiv\frac{1}{p_i^2-m_i^2+i\delta}$ can be parameterized by

\begin{equation*}
\frac{1}{D_i^{\lambda_i+1}}=\frac{e^{-\frac{\lambda_i+1}{2}i\pi}}{\Gamma(\lambda_i+1)}\int_0^{\infty}dx_i~e^{ix_iD_i}x_i^{\lambda_i},\qquad\text{Im}\{D_i\}>0.
\end{equation*}

\noindent A cut propagator can be parametrized similarly:

\begin{equation*}
-2\pi i\delta(D_i)=e^{-\frac{1}{2}\pi i}\int_{-\infty}^{\infty}dx_i~e^{ix_iD_i}.
\end{equation*}

\noindent By virtue of the identity

\begin{equation*}
l_{i_1}^{\mu_1}l_{i_2}^{\mu_2}\cdots l_{i_m}^{\mu_m}=\frac{i(-1)^m}{\Gamma(m+1)}\left[\frac{\partial}{\partial p_{i_1,\mu_1}}\frac{\partial}{\partial p_{i_2,\mu_2}}\cdots\frac{\partial}{\partial p_{i_m,\mu_m}}\int_0^\infty dy\exp\left[-iy\left(1+\sum_{i=1}^{L}p_i\cdot l_i\right)\right]\right]_{p_i^\mu=0},
\end{equation*}

\noindent a $d$ dimensional $L$-loop rank $m$ tensor integral with $n$ propagators can be generated by

\begin{equation}\label{eq:app:tmp2}
\begin{split}
\mathcal{M}^{\mu_1\mu_2\cdots\mu_m}(d,\lambda_1,\lambda_2,\cdots, \lambda_n)\equiv&\pi^{-Ld/2}\int d^dl_1d^dl_2\cdots d^dl_L\frac{l_{i_1}^{\mu_1}l_{i_2}^{\mu_2}\cdots l_{i_m}^{\mu_m}}{D_1^{\lambda_1+1}D_2^{\lambda_2+1}\cdots D_n^{\lambda_n+1}}\\
=&\frac{(-1)^m}{\Gamma(m+1)}\left[\frac{\partial}{\partial p_{i_1,\mu_1}}\frac{\partial}{\partial p_{i_2,\mu_2}}\cdots\frac{\partial}{\partial p_{i_m,\mu_m}}\mathcal{M}_p(d,\lambda_1,\lambda_2,\cdots, \lambda_n)\right]_{p_i^\mu=0},
\end{split}
\end{equation}

\noindent where $d_0$ is the real space-time dimension, and the generator

\begin{equation*}
\begin{split}
\mathcal{M}_p(d,\lambda_1,\lambda_2,\cdots, \lambda_n)\equiv&i\pi^{-Ld/2}\int dy\prod_{i=1}^n\left(dx_i\frac{x_i^{\lambda_i}e^{-\frac{i\pi}{2}(\lambda_i+1)}}{\Gamma(\lambda_i+1)}\right)\int \prod_{j=1}^Ld^dl_j\exp\left[i\sum_{i=1}^nx_iD_i-iy\left(1+\sum_{i=1}^{L}p_i\cdot l_i\right)\right]\\
\equiv&i\pi^{-Ld/2}\int dy\prod_{i=1}^n\left(dx_i\frac{x_i^{\lambda_i}e^{-\frac{i\pi}{2}(\lambda_i+1)}}{\Gamma(\lambda_i+1)}\right)\int \prod_{j=1}^Ld^dl_j\exp\left[i\left(\sum_{i,j=1}^LA_{ij}l_i\cdot l_j+2\sum_{i=1}^LB_i\cdot l_i+C\right)\right]\\
=&s_g^{-L/2}e^{\frac{i\pi}{2}\left(\frac{Ld}{2}+1\right)}\int dy\prod_{i=1}^n\left(dx_i\frac{x_i^{\lambda_i}e^{-\frac{i\pi}{2}(\lambda_i+1)}}{\Gamma(\lambda_i+1)}\right)\det(A)^{-\frac{d}{2}}\exp\left[i\left(C-\sum_{i,j}^L(A^{-1})_{ij}B_i\cdot B_j\right)\right].
\end{split}
\end{equation*}

\noindent Here $s_g$ is the determinant of the dimensionally regularized spacetime metric. For instance, in four dimensional Minkowski spacetime, we have $s_g=e^{i\pi(d-1)}$, and in four dimensional Euclidean space, we have $s_g=e^{i\pi(d-4)}$ (instead of $1$). To simplify the above integral, we insert a trivial integral $\int_0^\infty d\alpha(\alpha-E(x))$ into it, where $E(x)$ is a positive homogeneous function of $x_i$'s of degree $1$. In this paper, we choose $E(x)=\sum_i\left|x_i\right|$, where the sum in the delta function runs over any nontrivial subset of $\{x_1,x_2,\cdots x_n\}$. It should be noticed that $E(x)$ need not to be linear in $x$'s. Rescaling the variables of integration by $x_i\to\alpha x_i$, and $y\to\alpha y$, and integrating over $\alpha$, we get

\begin{equation}\label{eq:app:tmp5}
\begin{split}
\mathcal{M}_p(d,\lambda_1,\lambda_2,\cdots, \lambda_n)=&s_g^{-L/2}e^{i\pi\lambda_f}\frac{\Gamma(1-\lambda_f)}{\prod_{i=1}^n\Gamma(\lambda_i+1)}\int dy\prod_{i=1}^n\left(dx_ix_i^{\lambda_i}\right)\delta(1-E(x))U^{1-\frac{d}{2}-\lambda_f}F(p,y)^{\lambda_f-1}\\
=&s_g^{-L/2}e^{i\pi\lambda_f}\frac{\Gamma(\frac{d}{2})}{\Gamma(\lambda_f+\frac{d}{2}-1)\prod_{i=1}^n\Gamma(\lambda_i+1)}\\
&\times\int dy\prod_{i=1}^{n+1}dx_i\delta(1-E(x))\mathcal{F}(p,y)^{-\frac{d}{2}}x_{n+1}^{\lambda_f+\frac{d}{2}-2}\prod_{i=1}^nx_i^{\lambda_i}\\
\equiv&s_g^{-L/2}e^{i\pi\lambda_f}\frac{\Gamma(\frac{d}{2})}{\Gamma(\lambda_{n+1})\prod_{i=1}^{n}\Gamma(\lambda_i+1)}\int dyd\Pi^{(n+1)}\mathcal{F}(p,y)^{-\frac{d}{2}}x_{n+1}^{\lambda_{n+1}-1}\prod_{i=1}^nx_i^{\lambda_i},
\end{split}
\end{equation}

\noindent where $\lambda_f\equiv\frac{1}{2}dL-n-\sum_{i=1}^n\lambda_i$, $U\equiv\det(A)$, $F(p,y)\equiv U\left(\sum_{i,j}^L(A^{-1})_{ij}B_i\cdot B_j-C\right)$, $\mathcal{F}(p,y)\equiv Ux_{n+1}+F(p,y)$, and $d\Pi^{(n)}$ is the one defined in \cref{eq:FeynmInt}.

Generally, $\mathcal{F}(p,y)$ is of the form

\begin{equation}\label{eq:app:tmp1}
\begin{split}
\mathcal{F}(p,y)=&\mathcal{F}(0,0)+yU-yU\sum_{i,j=1}^L\left[(A^{-1})_{ij}B_i\cdot p_j\right]_{y=0}+\frac{1}{4}y^2U\sum_{i,j=1}^L\left[(A^{-1})_{ij}p_i\cdot p_j\right]_{y=0}\\
\equiv&\mathcal{F}(0,0)+yU+y\sum_{i=1}^Lb_i\cdot p_i+\sum_{i,j}^Lc_{ij}y^2p_i\cdot p_j,
\end{split}
\end{equation}

\noindent where $b$ and $c$ are polynomials of $x$'s. That is $b_i^\mu=b_i^\mu(x_1,x_2,\cdots,x_n)$, and $c_{ij}=c_{ij}(x_1,x_2,\cdots,x_n)$. Setting $p^\mu=0$, the integration over $y$ in \cref{eq:app:tmp5} can easily be carried out by shifting the variable $x_{n+1}\to x_{n+1}-y$.

\begin{equation}
\begin{split}
\mathcal{M}_0(d,\lambda_1,\lambda_2,\cdots, \lambda_n)=&s_g^{-L/2}e^{i\pi\lambda_f}\frac{\Gamma(\frac{d}{2})}{\Gamma(\lambda_{n+1})\prod_{i=1}^{n}\Gamma(\lambda_i+1)}\\
&\times\int_0^\infty dx_{n+1}\int_0^{x_{n+1}}dy\int d\Pi^{(n)}\mathcal{F}(0,0)^{-\frac{d}{2}}(x_{n+1}-y)^{\lambda_{n+1}-1}\prod_{i=1}^nx_i^{\lambda_i}\\
=&s_g^{-L/2}e^{i\pi\lambda_f}\frac{\Gamma(\frac{d}{2})}{\prod_{i=1}^{n+1}\Gamma(\lambda_i+1)}\int d\Pi^{(n+1)}\mathcal{F}(0,0)^{-\frac{d}{2}}\prod_{i=1}^{n+1}x_i^{\lambda_i}\\
\equiv&s_g^{-L/2}e^{i\pi\lambda_f}\int d\Pi^{(n+1)}\mathcal{I}(\lambda_0,\lambda_1,\cdots,\lambda_n)\\
\equiv&s_g^{-L/2}e^{i\pi\lambda_f}I(\lambda_0,\lambda_1,\cdots,\lambda_n)
\end{split}
\end{equation}

\noindent Similarly we have:

\begin{equation}\label{eq:app:tmp3}
\frac{\Gamma(\frac{d}{2})}{\Gamma(\lambda_{n+1}-\delta_y)\prod_{i=1}^{n}\Gamma(\lambda_i+1)}\int dyd\Pi^{(n+1)}\mathcal{F}(0,y)^{-\frac{d}{2}}y^{\delta_y}x_{n+1}^{\lambda_{n+1}-\delta_y-1}\prod_{i=1}^nx_i^{\lambda_i}=\Gamma(\delta_y+1)I(\lambda_0,\lambda_1,\cdots,\lambda_n).
\end{equation}

\noindent And obviously we have

\begin{equation}\label{eq:app:tmp4}
\begin{split}
&\frac{\Gamma(\frac{d}{2})}{\prod_{i=1}^{n+1}\Gamma(\lambda_i+1)}\int d\Pi^{(n)}x_i^{\delta_i}\mathcal{F}(0,0)^{-\frac{d}{2}}\prod_{j=1}^{n+1}x_j^{\lambda_j}\\
=&\prod_{j=1}^{\delta_i}(\lambda_i+j)I(d,\lambda_1,\lambda_2,\cdots,\lambda_i+\delta_i,\cdots,\lambda_n)\\
=&R_i^{\delta_i}I(-\frac{d}{2},\lambda_1,\lambda_2,\cdots,\lambda_n),
\end{split}
\end{equation}

\noindent where $R_i$ is an operator applying on $I$ such that $R_iI(\lambda_0,\lambda_1,\cdots,\lambda_i,\cdots,\lambda_n)\equiv(\lambda_i+1)I(\lambda_0,\lambda_1,\cdots,\lambda_i+1,\cdots,\lambda_n)$. Similarly we define $D_iI(\lambda_0,\lambda_1,\cdots,\lambda_i,\cdots,\lambda_n)\equiv I(\lambda_0,\lambda_1,\cdots,\lambda_i-1,\cdots,\lambda_n)$. We define the operator

\begin{equation}
P_{i,\mu}\equiv-\frac{\partial}{\partial p_i^\mu}+\Big[b_{i,\mu}(R_1,R_2,\cdots,R_n)+2\sum_{j=1}^Lp_{j,\mu}c_{ij}(R_1,R_2,\cdots,R_n)\Big]D_0,
\end{equation}

\noindent where $b$ and $c$ are defined in \cref{eq:app:tmp1}. Then by virtue of \cref{eq:app:tmp2,eq:app:tmp5,eq:app:tmp1,eq:app:tmp3,eq:app:tmp4}, it's easy to see that

\begin{equation}\label{eq:app:ParTensInt}
\mathcal{M}^{\mu_1\mu_2\cdots\mu_m}(d,\lambda_1,\lambda_2,\cdots, \lambda_n)=s_g^{-L/2}e^{i\pi\lambda_f}\left[P_{i_1}^{\mu_1}P_{i_2}^{\mu_2}\cdots P_{i_m}^{\mu_m}I(-\frac{d}{2},\lambda_1,\lambda_2,\cdots,\lambda_n)\right]_{p^\mu=0}.
\end{equation}

\noindent Notice that the factor of $\frac{1}{\Gamma(m+1)}$ in \cref{eq:app:tmp2} is canceled by the factor $\Gamma(\delta_y+1)$ in \cref{eq:app:tmp3}, because after applying the differential operators to $\mathcal{M}_p$ and putting $p^\mu=0$, the degree in $y$ of the integrand in \cref{eq:app:tmp5} is exactly $m$.

\section{Derivation of \cref{eq:IBP1}}\label{app:DerIdent}

By virtue of the homogeneity of the integrand $\mathcal{I}^{(-n-1)}$ in \cref{eq:FeynmInt2}, rescaling of the variables of integration leads to

\begin{equation*}
\begin{split}
&\int d\Pi^{(n+1)}\mathcal{F}^{\lambda_0}\prod_{i=1}^{n+1}x_i^{\lambda_i}\\
=&\int dx_1dx_2\cdots dx_{n+1}\delta\left(x_{n+1}/\alpha-1\right)\mathcal{F}^{\lambda_0}\prod_{i=1}^{n+1}x_i^{\lambda_i}\\
=&\int dx_1dx_2\cdots dx_n\left.\mathcal{F}^{\lambda_0}\right|_{x_{n+1}=\alpha}\alpha^{\lambda_{n+1}+1}\prod_{i=1}^{n}x_i^{\lambda_i}\\
\equiv&G(\alpha),\quad \lambda\notin\mathbb{Z}^-.
\end{split}
\end{equation*}

\noindent Obviously $G(\alpha)$ should be independent of $\alpha$. Thus we have

\begin{equation*}
\begin{split}
0=&\alpha\frac{\partial G(\alpha)}{\partial\alpha}\\
=&\int dx_1dx_2\cdots dx_n\alpha\frac{\partial}{\partial \alpha}\left[\left.\mathcal{F}^{\lambda_0}\right|_{x_{n+1}=\alpha}\alpha^{\lambda_{n+1}+1}\prod_{i=1}^{n}x_i^{\lambda_i}\right]\\
=&\int dx_1dx_2\cdots dx_{n+1}\delta(x_{n+1}/\alpha-1)\frac{\partial}{\partial x_{n+1}}\left[\mathcal{F}^{\lambda_0}x_{n+1}^{\lambda_{n+1}+1}\prod_{i=1}^{n}x_i^{\lambda_i}\right]\\
=&\int d\Pi^{(n+1)}\frac{\partial}{\partial x_{n+1}}\left[\mathcal{F}^{\lambda_0}x_{n+1}^{\lambda_{n+1}+1}\prod_{i=1}^{n}x_i^{\lambda_i}\right],\quad\lambda\notin\mathbb{Z}^-.
\end{split}
\end{equation*}

\noindent Since $\lambda\notin\mathbb{Z}^-$, $\lambda_{n+1}+1\neq0$. Replacing $\lambda_{n+1}+1$ by $\lambda_{n+1}$, we get

\begin{equation*}
0=\int d\Pi^{(n+1)}\frac{\partial}{\partial x_{n+1}}\mathcal{I}^{(-n)},\quad\lambda_{n+1}\neq0,~\lambda\notin\mathbb{Z}^-.
\end{equation*}

\noindent Similarly, for a general $x_i$, we have

\begin{equation}
0=\int d\Pi^{(n+1)}\frac{\partial}{\partial x_i}\mathcal{I}^{(-n)},\quad\lambda_i\neq0,~\lambda\notin\mathbb{Z}^-.
\end{equation}

\noindent It's easy to prove that this equation still holds in the case where the domain of integration of $x_i$ is $(-\infty,\infty)$.

We consider the limit $\lambda_i\to0$. By using the formula $\frac{1}{x^{1-\lambda}}=\frac{1}{\lambda}\delta(x)+\mathcal{O}(\lambda^0)$ \cite{Actis:2004bp}, we get

\begin{equation*}
\begin{split}
0=&\lim_{\lambda_i\to0}\int d\Pi^{(n+1)}\frac{\partial}{\partial x_i}\left[\mathcal{F}^{\lambda_0}x_i^{\lambda_i}\prod_{j\neq i}^{n+1}x_j^{\lambda_j}\right]\\
=&\int d\Pi^{(n+1)}\frac{\partial}{\partial x_i}\left[\mathcal{F}^{\lambda_0}\prod_{j\neq i}^{n+1}x_j^{\lambda_j}\right]+\int d\Pi^{(n)}\left.\mathcal{F}^{\lambda_0}\right|_{x_i=0}\prod_{j\neq i}^{n+1}x_j^{\lambda_j},\quad\lambda\notin\mathbb{Z}^-.
\end{split}
\end{equation*}

\noindent Then we have

\begin{equation}
0=\int d\Pi^{(n+1)}\frac{\partial}{\partial x_i}\mathcal{I}^{(-n)}+\int d\Pi^{(n)}\left.\mathcal{I}^{(-n)}\right|_{x_i=0},\quad\lambda_i=0,~\lambda\notin\mathbb{Z}^-.
\end{equation}


\begin{thebibliography}{99}

\bibitem{Tkachov:1981wb} 
  F.~V.~Tkachov,
  Phys.\ Lett.\  {\bf 100B}, 65 (1981).

\bibitem{Chetyrkin:1981qh} 
  K.~G.~Chetyrkin and F.~V.~Tkachov,
  Nucl.\ Phys.\ B {\bf 192}, 159 (1981).

\bibitem{Baikov:1996iu} 
  P.~A.~Baikov,
  Nucl.\ Instrum.\ Meth.\ A {\bf 389}, 347 (1997).

\bibitem{Tarasov:1998nx} 
  O.~V.~Tarasov,
  Acta Phys.\ Polon.\ B {\bf 29}, 2655 (1998).

\bibitem{Laporta:2001dd} 
  S.~Laporta,
  Int.\ J.\ Mod.\ Phys.\ A {\bf 15}, 5087 (2000).

\bibitem{Smirnov:2005ky} 
  A.~V.~Smirnov and V.~A.~Smirnov,
  JHEP {\bf 0601}, 001 (2006).

\bibitem{Lee:2008tj} 
  R.~N.~Lee,
  JHEP {\bf 0807}, 031 (2008).

\bibitem{Kant:2013vta} 
  P.~Kant,
  Comput.\ Phys.\ Commun.\  {\bf 185}, 1473 (2014).

\bibitem{vonManteuffel:2014ixa} 
  A.~von Manteuffel and R.~M.~Schabinger,
  Phys.\ Lett.\ B {\bf 744}, 101 (2015).

\bibitem{Anastasiou:2004vj} 
  C.~Anastasiou and A.~Lazopoulos,
  JHEP {\bf 0407}, 046 (2004).

\bibitem{Smirnov:2008iw} 
  A.~V.~Smirnov,
  JHEP {\bf 0810}, 107 (2008).

\bibitem{Studerus:2009ye} 
  C.~Studerus,
  Comput.\ Phys.\ Commun.\  {\bf 181}, 1293 (2010).

\bibitem{Lee:2012cn} 
  R.~N.~Lee,
  arXiv:1212.2685 [hep-ph].
  
\bibitem{vonManteuffel:2016xki} 
  A.~von Manteuffel and R.~M.~Schabinger,
  Phys.\ Rev.\ D {\bf 95}, 034030 (2017).

\bibitem{Maierhoefer:2017hyi} 
  P.~Maierhöfer, J.~Usovitsch and P.~Uwer,
  Comput.\ Phys.\ Commun.\  {\bf 230}, 99 (2018).


\bibitem{Smirnov:2019qkx} 
  A.~V.~Smirnov and F.~S.~Chuharev,
  arXiv:1901.07808 [hep-ph].

\bibitem{Gehrmann:1999as} 
  T.~Gehrmann and E.~Remiddi,
  Nucl.\ Phys.\ B {\bf 580}, 485 (2000).

\bibitem{Lee:2014tja} 
  R.~N.~Lee,
  arXiv:1405.5616 [hep-ph].
  
\bibitem{Lee:2013hzt} 
  R.~N.~Lee and A.~A.~Pomeransky,
  JHEP {\bf 1311}, 165 (2013)

\bibitem{Chen:2019fzm} 
  W.~Chen,
  arXiv:1912.08606 [hep-ph].

\bibitem{Feynman:1949zx} 
  R.~P.~Feynman,
  Phys.\ Rev.\  {\bf 76}, 769 (1949).

\bibitem{Nakanishi:1957aa}
  N.~Nakanishi,
  Progr.\ Theor.\ Phys {\bf 17}, 401 (1957).

\bibitem{Symanzik:1958aa}
  K.~Symanzik,
  Progr.\ Theor.\ Phys {\bf 20}, 690 (1958).

\bibitem{Tarasov:1996br} 
  O.~V.~Tarasov,
  Phys.\ Rev.\ D {\bf 54}, 6479 (1996).

\bibitem{Bitoun:2017nre} 
  T.~Bitoun, C.~Bogner, R.~P.~Klausen and E.~Panzer,
  Lett.\ Math.\ Phys.\  {\bf 109}, no. 3, 497 (2019)

\bibitem{Larsen:2015ped} 
  K.~J.~Larsen and Y.~Zhang,
  Phys.\ Rev.\ D {\bf 93}, no. 4, 041701 (2016)

\bibitem{Boehm:2017wjc} 
  J.~B\"ohm, A.~Georgoudis, K.~J.~Larsen, M.~Schulze and Y.~Zhang,
  Phys.\ Rev.\ D {\bf 98}, no. 2, 025023 (2018)

\bibitem{Boehm:2018fpv} 
  J.~B\"ohm, A.~Georgoudis, K.~J.~Larsen, H.~Sch\"onemann and Y.~Zhang,
  JHEP {\bf 1809}, 024 (2018)

\bibitem{Lee:2013mka} 
  R.~N.~Lee,
  J.\ Phys.\ Conf.\ Ser.\  {\bf 523}, 012059 (2014).

\bibitem{Smirnov:2008py} 
  A.~V.~Smirnov and M.~N.~Tentyukov,
  Comput.\ Phys.\ Commun.\  {\bf 180}, 735 (2009).

\bibitem{Actis:2004bp} 
  S.~Actis, A.~Ferroglia, G.~Passarino, M.~Passera and S.~Uccirati,
  Nucl.\ Phys.\ B {\bf 703}, 3 (2004)
  
\end{thebibliography}
\end{document}